\documentstyle[preprint,prc,aps]{revtex}

\begin{document}

\title{The Rarita-Schwinger spin-${3\over2}$ equation in a nonuniform,
central potential}

\author{P.~G.~Blunden, Y.~Okuhara and A.~S.~Raskin}

\address{Department of Physics, University of Manitoba, Winnipeg,
Manitoba, Canada\ \ R3T~2N2}

\maketitle

\begin{abstract}

The equations of motion for a massive spin-${3\over2}$ Rarita-Schwinger
field in a finite-range, central, Lorentz scalar potential are developed. 
It is shown that the resulting density may not be everywhere positive
definite.

\end{abstract}

\section{Introduction}

For many years there has been interest in studying spin-${3\over2}$
particles in as much detail as has been done for spin-${1\over2}$
particles.  In the latter case, the Dirac equation has been shown to
provide a reasonable starting basis for such studies \cite{Serot_Walecka},
although of course the formalism of quantum field theory is needed for any
detailed consideration.  Jasiak and Szymacha \cite{Jasiak} have attempted
to use the Rarita-Schwinger equations for spin-${3\over2}$ particles in
order to investigate the $\Delta$ resonance in the bag model. However,
they restricted themselves to solving only a subset of the equations, and
made no attempt to verify whether or not the wave function that they
calculated was physically meaningful. In this paper, we address this
question by explicitly considering the most general form for the wave
function.

Attempts to quantize the spin-${3\over2}$ Rarita-Schwinger field in the
presence of interactions of scalar and vector type are beset with problems
\cite{Hagen}.  For example, it is known that the Rarita-Schwinger field
has noncausal modes of propagation \cite{Velo}.  Johnson and Sudarshan
\cite{Johnson} found that for vector interactions, the canonical
anti-commutator is not positive definite at all spacetime points.  Here we
will restrict our considerations to the classical field equations. 

Our work here is motivated by relativistic QHD models of Walecka and
collaborators \cite{Serot_Walecka,HS}, which consider nucleons interacting
with mean-field scalar and vector potentials by the Dirac equation.  Our
initial purpose in undertaking this work was to describe the $\Delta$ by
an analogous relativistic equation.  This is clearly important in order to
obtain a full treatment of the electromagnetic response of the nucleus,
particularly for deep inelastic electron scattering in the quasi-elastic
regime and beyond \cite{Wehrberger}.  An alternative approach, taken by
Lin and Serot \cite{Lin_Serot}, is to treat the $\Delta$ dynamically as a
$\pi-N$ resonance, thus avoiding the need for an explicit equation for the
$\Delta$.  Here we treat the $\Delta$ as an additional degree of freedom,
described by the relativistic spin-${3\over2}$ Rarita-Schwinger equation. 
This is analogous to the non-relativistic $\Delta$-hole models. 
 
\section{Formalism for Nonuniform, Central Scalar Potentials}

The free-particle Rarita-Schwinger equations
$(i\,\rlap{/}\partial-m)\Psi^\mu=0$, and the auxiliary condition
$\gamma_\mu \Psi^\mu=0$, can both be derived from the Lagrangian
\begin{equation}
{\cal L}_{RS} = \overline{\Psi}_\mu(\rlap{/}\partial+i\,m)
\Psi^\mu-{1\over3}\overline{\Psi}_\mu(
\partial^\mu\Phi+\gamma^\mu\partial\cdot\Psi)+{1\over3}\overline{\Psi}_\mu
\gamma^\mu(\rlap{/}\partial-i\,m)\Phi,\label{lscr}
\end{equation}
where $\rlap{/}\partial\equiv\gamma_\mu \partial^\mu$,
$\Phi\equiv\gamma_\mu \Psi^\mu$, and $\partial\cdot\Psi\equiv\partial_\mu
\Psi^\mu$. The Euler-Lagrange equations of motion corresponding to ${\cal
L}_{RS}$ are given by
\begin{equation}
(\rlap{/}\partial+i\,m)\Psi^\mu-{1\over3}(\partial^\mu\Phi+\gamma^\mu\partial
\cdot\Psi) +{1\over3}\gamma^\mu(\rlap{/}\partial-i\,m)\Phi=0.\label{rs}
\end{equation}
Contracting this result with $\gamma_\mu$ and $\partial_\mu$ results in
the subsidiary conditions $\Phi=0$, and $\partial\cdot\Psi=0$.  Hence
Eq.~(\ref{rs}) reduces to the usual Rarita-Schwinger equations.  Note that
Eq.~(\ref{lscr}) is not the most general form for ${\cal L}_{RS}$. In
fact, there is a two-parameter Lagrangian that leads to exactly the same
free equations of motion regardless of the values of the parameters. 
 
The most general form of bilinear, non-derivative couplings of a massive
Rarita-Schwinger field was investigated by Hagen and Singh \cite{Hagen}. 
Here we are interested in the simplest case of a Lorentz scalar potential. 
In order to introduce this interaction potential into the Rarita-Schwinger
formalism, we follow the approach used for the Dirac equation (and applied
to the spin-${3\over2}$ case by Jasiak and Szymacha \cite{Jasiak})  and
replace the constant mass $m=m_0$ in the Lagrangian and resulting
Euler-Lagrange equations by a mass $m(r)=m_0-U(r)$, which is a spherically
symmetric function of radius $r$.  This approach is consistent with that
of Hagen and Singh \cite{Hagen}.  The scalar potential $U(r)$ is of finite
range, and vanishes at sufficiently large values of $r$. 

The subsidiary conditions that were derived from Eq.~(\ref{rs}) now become
\begin{eqnarray}
\partial\cdot\Psi+{1\over m}{dm\over dr}\,\hat{\bbox{r}}\cdot\bbox{\Psi}&=0&,
\label{sub1}\\
2\partial\cdot\Psi-i\,m\Phi&=0&.\label{sub2}
\end{eqnarray}
The Euler-Lagrange equations can be written in the forms:
\begin{eqnarray}
(\rlap{/}\partial+i\,m)(\Psi^0-{1\over3}\gamma^0\Phi)+{1\over6}(2\partial^0-i\,m
\gamma^0)\Phi&=0&,\label{el1}\\
\hat{\bbox{r}}\cdot(\rlap{/}\partial+i\,m)\bbox{\Psi}+
{1\over3}{\partial\Phi\over\partial r}-{1\over3}\gamma^r
\partial\cdot\Psi+{1\over3}\gamma^r(\rlap{/}\partial-i\,m)\Phi&=0&,\label{el2}
\end{eqnarray}
where $\gamma^r\equiv\hat{\bbox{r}}\cdot\bbox{\gamma}$.  These two
equations, along with the two subsidiary conditions (\ref{sub1})  and
(\ref{sub2}), replace the original four Rarita-Schwinger equations. 

Using the usual Dirac representation of the gamma matrices $\gamma^\mu$,
it follows that
$\rlap{/}\partial = \gamma^0\partial^0+\gamma^r\left({\partial\ 
\over\partial r}+ {1+\hat{\kappa}\over r}\right)$, where
$\hat{\kappa}\equiv-(1+\bbox{\sigma}\cdot\bbox{L})$.
Then Eq.~(\ref{el2}) reduces to 
\begin{eqnarray}
(\gamma^0\partial^0+i\,m)(m\,\Psi^r)&+&\left({\partial\ \over\partial r}+
{2-\hat{\kappa}\over r}\right)(m\,\gamma^r\Psi^r)\nonumber\\
&-&{1\over3}\left(\gamma^0\gamma^r\partial^0-{1-\hat{\kappa}\over r}\right)
(m\,\Phi)-
{m\over r}\gamma^0\left(\Psi^0-{1\over3}\gamma^0\Phi\right)=0,\label{el2red}
\end{eqnarray}
where $\Psi^r\equiv\hat{\bbox{r}}\cdot\bbox{\Psi}$.
 
In order to solve these equations, we seek normal-mode solutions for
$\Psi^\mu(\bbox{r},t)$ in their most general form consistent with
conservation of angular momentum.  For the time-like component, we have
\begin{equation}
\Psi^0(\bbox{r},t)={\rm e}^{-iEt}\,{1\over r}\,\left(
\begin{array}{c}
G_0(r)\,i\,Y_\ell^j(\hat{\bbox{r}})\\
-F_0(r)\,Y_{\ell^\prime}^j(\hat{\bbox{r}})
\end{array}
\right),\label{eq6}
\end{equation}
where $Y_\ell^j(\hat{\bbox{r}})\equiv
[Y_\ell(\hat{\bbox{r}})\otimes\chi]^j$, $\chi$ is the spin-${1\over2}$
spinor, and $\ell^\prime\equiv2j-\ell$ is the other value of $\ell$
possible for the given value of the angular momentum quantum number $j$. 
For the space-like component, we have the superposition
\begin{equation}
\bbox{\Psi}(\bbox{r},t)={\rm e}^{-iEt}\,{1\over r}\,\left(
\begin{array}{c}
G_1(r)\,\bbox{Y}^j_{\ell_1 {3\over2}}(\hat{\bbox{r}})+
G_2(r)\,\bbox{Y}^j_{\ell_2 {3\over2}}(\hat{\bbox{r}})+
G_3(r)\,\bbox{Y}^j_{\ell^\prime{1\over2}}(\hat{\bbox{r}})\\
F_1(r)\,i\,\bbox{Y}^j_{\tilde\ell_1 {3\over2}}(\hat{\bbox{r}})+
F_2(r)\,i\,\bbox{Y}^j_{\tilde\ell_2 {3\over2}}(\hat{\bbox{r}})+
F_3(r)\,i\,\bbox{Y}^j_{\ell {1\over2}}(\hat{\bbox{r}})
\end{array}
\right),\label{eq7}
\end{equation}
where $\bbox{Y}_{\ell
s}^j(\hat{\bbox{r}})\equiv[Y_\ell(\hat{\bbox{r}})\otimes [\chi\otimes
\bbox{e}_1]_s]^j$, with $\bbox{e}_1$ the unit vector.  The values of
$\ell_1, \ell_2, \tilde\ell_1$ and $\tilde\ell_2$ in Eq.~(\ref{eq7}) are
constrained by angular momentum considerations ($\bbox{\ell}_i=\bbox{j}+
{\bf{3\over2}}$), and by the Rarita-Schwinger equations themselves, as
will be seen shortly. 
 
Before trying to solve Eqs.~(\ref{el1}) and (\ref{el2red}), we first
consider several identities for a vector field field $\bbox{V}_{\ell
s}^j(\bbox{r})\equiv f_\ell(r)  \bbox{Y}_{\ell s}^j(\hat{\bbox{r}})$.
These are
\begin{mathletters}
\begin{eqnarray}
\hat{\bbox{r}}\cdot\bbox{V}_{\ell s}^j(\bbox{r})&=&\sum_{\ell^\prime}
\alpha_{j\ell s}^{\ell^\prime}\,Y_{\ell^\prime}^j(\hat{\bbox{r}})\,f_\ell(r),
\label{eq8a}\\
\bbox{\sigma}\cdot\bbox{V}_{\ell s}^j(\bbox{r})&=&\sqrt{3}\,Y_\ell^j
(\hat{\bbox{r}})\,\delta_{s,{1\over2}}
\ f_\ell(r),\label{eq8b}\\
\bbox{\nabla}\cdot\bbox{V}_{\ell s}^j(\bbox{r})&=&\sum_{\ell^\prime}
\alpha_{j\ell s}^{\ell^\prime}\,Y_{\ell^\prime}^j(\hat{\bbox{r}})
\left({d\ \over dr}+{\beta_{\ell^\prime\ell}\over r}\right)f_\ell(r),
\label{eq8c}
\end{eqnarray}
\end{mathletters}
where
\begin{mathletters}
\begin{eqnarray}
\beta_{\ell^\prime\ell}&\equiv&
\left\{
\begin{array}{cl}
-\ell,&\mbox{if\ $\ell^\prime=\ell+1$;}\\
(\ell+1),&\mbox{if\ $\ell^\prime=\ell-1$;}
\end{array}\right.\label{eq9a}\\
\alpha_{j\ell s}^{\ell^\prime}&\equiv&[s][\ell][\ell^\prime]
\ (-1)^{j+s-1}
\biggl(
\begin{array}{ccc}
\ell&1&\ell^\prime\\
\noalign{\vskip -35pt}\\
0&0&0
\end{array}\biggr)
\biggl\{
\begin{array}{ccc}
\ell&1&\ell^\prime\\
\noalign{\vskip -35pt}\\
{1\over2}&j&s
\end{array}\biggr\},
\label{eq9b}
\end{eqnarray}
\end{mathletters}
with $[j]\equiv\sqrt{2j+1}$. It then follows from putting Eqs.~(\ref{eq6})
and (\ref{eq7}) into Eq.~(\ref{sub2}) that
\begin{equation}
\Phi={\rm e}^{-iEt}\,{1\over r}\,\left(
\begin{array}{c}
(G_0(r)-\sqrt{3} F_3(r))\,i\,Y^j_\ell\\
(F_0(r)+\sqrt{3} G_3(r))\,Y^j_{\ell^\prime}
\end{array}\right).\label{eq10}
\end{equation}
Hence Eq.~(\ref{el1}) reduces to the set
\begin{mathletters}
\label{eq11}
\begin{eqnarray}
(2E-m)G_0(r)-m\sqrt{3} F_3(r)+{2\over3}\left({d\ \over dr}-{\kappa\over r}
\right)(2F_0(r)-\sqrt{3} G_3(r))&=&0,\\
-(2E+m)F_0(r)+m\sqrt{3} G_3(r)+{2\over3}\left({d\ \over dr}+{\kappa\over r}
\right)(2G_0(r)+\sqrt{3} F_3(r))&=&0,
\end{eqnarray}
\end{mathletters}
with $\kappa=-(\ell+1)$ if $j=\ell+{1\over2}$, and $\kappa=\ell$ if
$j=\ell-{1\over2}$. 
 
For the other three sets of equations, it is useful to define the
following quantities: 
\begin{eqnarray}
\begin{array}{ccc}
\tilde{G}_1(r)\equiv {1\over2}\left[
\begin{array}{r}
\sqrt{2j-1\over j}\\
\sqrt{2j-1\over3(j+1)}
\end{array}\right]G_1(r),\qquad&
\tilde{G}_2(r)\equiv {1\over2}\left[
\begin{array}{r}
\sqrt{2j+3\over 3j}\\
\sqrt{2j+3\over(j+1)}
\end{array}\right]G_2(r),\qquad&\\
\noalign{\vskip 10pt}\cr
\tilde{F}_1(r)\equiv {1\over2}\left[
\begin{array}{r}
\sqrt{2j-1\over 3(j+1)}\\
\sqrt{2j-1\over3(j+1)}
\end{array}\right]F_1(r),\qquad&
\tilde{F}_2(r)\equiv {1\over2}\left[
\begin{array}{r}
\sqrt{2j+3\over j+1}\\
\sqrt{2j+3\over 3j}
\end{array}\right]F_2(r),\qquad&\\
\noalign{\vskip 10pt}\cr
\alpha\equiv\left[
\begin{array}{r}
-(j-{1\over2})\\
-(j+{1\over2})
\end{array}\right],\qquad&
\alpha^\prime\equiv\left[
\begin{array}{r}
-(j+{1\over2})\\
-(j-{1\over2})
\end{array}\right],\qquad&
\kappa\equiv\left[
\begin{array}{r}
-(j+{1\over2})\\
(j+{1\over2})
\end{array}\right],
\end{array}
\end{eqnarray}
where the upper (lower) component applies for $j=\ell+{1\over2}$
($j=\ell-{1\over2}$). We also introduce the following four functions:
\begin{mathletters}
\begin{eqnarray}
\tilde G(r)&\equiv&\tilde G_1(r)-\tilde G_2(r)-{1\over\sqrt{3}}G_3(r),\\
\tilde F(r)&\equiv&\tilde F_1(r)-\tilde F_2(r)-{1\over\sqrt{3}}F_3(r),\\
\tilde G^\prime(r)&\equiv&\alpha\tilde G_1(r)-(1-\alpha)
\tilde G_2(r)+{\kappa\over\sqrt{3}} G_3(r),\\
\tilde F^\prime(r)&\equiv&\alpha^\prime\tilde F_1(r)-(1-\alpha^\prime)\tilde F_2(r)-
{\kappa\over\sqrt{3}}F_3(r).
\end{eqnarray}
\end{mathletters}
Then Eq.~(\ref{sub1}), together with (\ref{sub2}), implies that
$\Phi={2i\over m^2} {dm\over dr}\Psi^r$, and so we find
\begin{mathletters}
\label{eq14}
\begin{eqnarray}
G_0(r)-\sqrt{3} F_3(r)&=&{2\over m^2}{dm\over dr}\tilde G(r),\\
F_0(r)+\sqrt{3} G_3(r)&=&-{2\over m^2}{dm\over dr}\tilde F(r).
\end{eqnarray}
\end{mathletters}
The angular momentum constraints for
$\ell=j-{1\over2}$ require $\ell^\prime=j+{1\over2}$, $\ell_1=j-{3\over2}$, 
$\ell_2=j+{1\over2}$, $\tilde\ell_1=j-{1\over2}$ and $\tilde\ell_2=
j+{3\over2}$.  Similarly, the constraints for
$\ell=j+{1\over2}$ require $\ell^\prime=j-{1\over2}$, $\ell_1=j-{1\over2}$, 
$\ell_2=j+{3\over2}$, $\tilde\ell_1=j-{3\over2}$ and $\tilde\ell_2=j+{1\over2}$.
Eq.~(\ref{sub2}) reduces to
\begin{mathletters}
\label{eq15}
\begin{eqnarray}
(2E+m)G_0(r)-m\sqrt{3} F_3(r)+2{d\tilde G(r) \over dr}+2{\tilde G^\prime(r)\over r}&=&0,\\
(2E-m)F_0(r)-m\sqrt{3} G_3(r)+2{d\tilde F(r)\over dr}+2{\tilde F^\prime(r)\over r}&=&0.
\end{eqnarray}
\end{mathletters}
Finally, Eq.~(\ref{el2red}) implies that
\begin{mathletters}
\label{eq16}
\begin{eqnarray}
-(E-m)(m\tilde G(r))&-&{m\over3r}(2G_0(r)+\sqrt{3} F_3(r))-{1\over3}E\,m
(F_0(r)+\sqrt{3} G_3(r)) \nonumber\\
&+&{m\over3r}(1-\kappa)(G_0(r)-\sqrt{3} F_3(r))
-\left({d\ \over dr}+{1-\kappa\over r}\right)(m\tilde F(r))=0,\\
-(E+m)(m\tilde F(r))&-&{m\over3r}(2F_0(r)-\sqrt{3} G_3(r))+{1\over3}E\,m
(G_0(r)-\sqrt{3} F_3(r)) \nonumber\\
&+&{m\over3r}(1+\kappa)(F_0(r)+\sqrt{3} G_3(r))
+\left({d\ \over dr}+{1+\kappa\over r}\right)(m\tilde G(r))=0.
\end{eqnarray}
\end{mathletters}
 
Thus we have derived eight equations (\ref{eq11}, \ref{eq14}--\ref{eq16}),
in eight unknown functions, that together constitute the Rarita-Schwinger
equations in a finite, spherically symmetric system.  Eqs.~(\ref{eq14})
can be used to eliminate $G_3(r)$ and $F_3(r)$, while Eqs.~(\ref{eq15})
give $\tilde G^\prime(r)$ and $\tilde F^\prime(r)$ in terms of $G_0(r)$,
$F_0(r)$, $\tilde G(r)$, and $\tilde F(r)$. We are then left with four
linear, first-order differential equations involving only four functions. 
This system of equations (with energy eigenvalue $E$) can be readily
solved using standard numerical techniques. 

\section{Probability Density}

We now turn our attention to the probability density for the system.  The
conserved current $J^\mu$ corresponding to the Lagrangian ${\cal L}_{RS}$
is given by
\begin{equation}
J^\mu=\overline\Psi_\alpha\gamma^\mu\Psi^\alpha-{1\over3}\left(\Phi^\dagger
\gamma^0\Psi^\mu+\overline\Psi^\mu\Phi\right)+{1\over3}\Phi^\dagger\gamma^0
\gamma^\mu\Phi.\label{eq17}
\end{equation}
Explicitly, we have that $\partial\cdot J=0$, if $\Psi^\mu$ satisfies the
Euler-Lagrange equations. Then $\rho\equiv J^0=\rho_0-
\bbox{\Psi}^\dagger\cdot \bbox{\Psi}$ can be considered to define the
probability density of the spin-${3\over2}$ particle, where
\begin{eqnarray}
\rho_0&=&{\Psi^0}^\dagger\Psi^0-{1\over3}(\Phi^\dagger\gamma^0\Psi^0+\Psi^{0
\dagger}\gamma^0\Phi)+{1\over3}\Phi^\dagger\Phi\nonumber\\
&=&{1\over r^2}|Y_\ell^j|^2\left\{{2\over3}G_0^2(r)+F_3^2(r)\right\}+
{1\over r^2}|Y_{\ell^\prime}^j|^2\left\{{2\over3}F_0^2(r)+G_3^2(r)\right\},
\end{eqnarray}
using the explicit forms for $\Psi^0$ and $\Phi$. The quantity
$\bbox{\Psi}^\dagger\cdot \bbox{\Psi}$
can be evaluated from Eq.~(\ref{eq7}), leading to
\begin{eqnarray}
\rho(r)&=&{1\over4\pi r^2}[j]^2
\sum_{{\cal L} {\rm even}}[{\cal L}]^2
\biggl(
\begin{array}{ccc}
j&j&{\cal L}\\
\noalign{\vskip -35pt}\\
m_j&-m_j&0
\end{array}\biggr)
P_{\cal L}(\cos\theta)(-1)^{m_j+1/2}\nonumber\\
&&\quad\times\Biggl[\ [\ell]^2
\biggl\{
\begin{array}{ccc}
{\cal L}&j&j\\
\noalign{\vskip -35pt}\\
{1\over2}&\ell&\ell
\end{array}\biggr\}
\biggl(
\begin{array}{ccc}
\ell&\ell&{\cal L}\\
\noalign{\vskip -35pt}\\
0&0&0
\end{array}\biggr)
{2\over3}G_0^2(r)\,+\, [\ell^\prime]^2
\biggl\{
\begin{array}{ccc}
{\cal L}&j&j\\
\noalign{\vskip -35pt}\\
{1\over2}&\ell^\prime&\ell^\prime
\end{array}\biggr\}
\biggl(
\begin{array}{ccc}
\ell^\prime&\ell^\prime&{\cal L}\\
\noalign{\vskip -35pt}\\
0&0&0
\end{array}\biggr)
{2\over3}F_0^2(r) \nonumber\\
&&\qquad +\,\sum_{i,k=1}^2\Biggl(
G_i(r)\,G_k(r)\,[\ell_i][\ell_k]
\biggl\{
\begin{array}{ccc}
{\cal L}&j&j\\
\noalign{\vskip -35pt}\\
{3\over2}&\ell_i&\ell_k
\end{array}\biggr\}
\biggl(
\begin{array}{ccc}
\ell_i&\ell_k&{\cal L}\\
\noalign{\vskip -35pt}\\
0&0&0
\end{array}\biggr)\nonumber\\
&&\qquad\qquad\qquad+\,F_i(r)\,F_k(r)\,[\tilde\ell_i][\tilde\ell_k]
\biggl\{
\begin{array}{ccc}
{\cal L}&j&j\\
\noalign{\vskip -35pt}\\
{3\over2}&\tilde\ell_i&\tilde\ell_k
\end{array}\biggr\}
\biggl(
\begin{array}{ccc}
\tilde\ell_i&\tilde\ell_k&{\cal L}\\
\noalign{\vskip -35pt}\\
0&0&0
\end{array}\biggr)
\Biggr)\Biggr],\\
&=&{1\over4\pi r^2}\left\{{2\over3}G_0^2(r)-G_1^2(r)-G_2^2(r)+
{2\over3}F_0^2(r)-F_1^2(r)-F_2^2(r)
\right\}\,+\,\mbox{\rm higher-${\cal L}$ terms}.\label{density}
\end{eqnarray}
For $j={1\over2}$, only the above ${\cal L}=0$ term is allowed. Similarly,
only this term contributes for $j={3\over2}$ at
$\cos\theta=\pm{1\over\sqrt{3}}$. 

Explicit numerical solution of the Rarita-Schwinger equations,
Eqs.~(\ref{eq11}) and (\ref{eq16}), as well as the dependent
Eqs.~(\ref{eq14}) and (\ref{eq15}), for $j={1\over2}$ ($\ell=0,1$) and
$j={3\over2}$ ($\ell=1,2$)  were undertaken for several physically
plausible potentials.  In particular, we examined the solutions for a
particle of mass 1230 MeV in potentials of nuclear range (a few fm.) and
depth (a few hundred MeV), corresponding to the $\Delta$ in a nuclear
environment.  Although reasonable values of the energy eigenvalue were
found, the wave functions typically caused the term in brace brackets in
Eq.~(\ref{density}) to change sign as $r$ varies.  This corresponds to the
existence of a negative probability density, since only the overall sign
of $J^\mu$ lacks any physical significance.  This indicates either that
the interpretation of $\rho(r)$ as being a probability density is invalid,
or that the Rarita-Schwinger formalism with a scalar potential is
intrinsically unphysical. In addition, our attempts to use the two degrees
of freedom inherent in the more general form for ${\cal L}_{RS}$ to see if
another choice of the parameters could lead to a $\rho(r)$ that is
explicitly positive definite were also unsuccessful. All parameter sets
led to the same qualitative behavior as was observed for the Lagrangian
of Eq.~(\ref{lscr}). 

The lack of a positive definite probability density suggests that many of
the problems that plague attempts to quantize spin-${3\over 2}$ fields are
also to be found in the classical field equations. 

\acknowledgements
This work was supported in part by the Natural Sciences and Engineering Research
Council of Canada.

\end{document}